Perspective

# Collective Intelligence and Neurodynamics: Functional Homologies

William Sulis
Collective Intelligence Laboratory,
Department of Psychology and Behavioural Neurosciences
McMaster University, Hamilton, Canada
sulisw@mcmaster.ca

Abstract

A deep understanding of the dynamics of the human nervous system requires the simultaneous study of multiple spatiotemporal scales from the level of neurotransmitters up to the level of human cultures. This is likely impossible for technical and ethical reasons. Piecemeal analysis provides some understanding of the dynamics at single levels, but this does not illuminate the interactions between levels which are, at the very least, of great importance clinically. It would be useful to have an accessible biological system which could serve as a proxy for the nervous system and from which useful insights might be obtained. Functional homologies between the nervous system and collective intelligence systems, in particular social insect colonies, are described. It is proposed that social insects colonies could serve as functional proxies for nervous systems. Thus a multiscale study of social insect colonies may provide insights into the dynamics of nervous systems..

Introduction

Psychoanalysis provided the first modern general theory of mind. Freud attempted to give it a rigorous scientific foundation, but neuroscience and psychology were in their infancy, and he was forced to draw on metaphors from the physical sciences which were not appropriate to the task. Remarkably, another general theory of mind had been formulated 2600 years ago but went unrecognized in the West until recent times. This is the theory of mind of Buddhist psychology [1] which is grounded, not in mechanistic concepts, but rather in ideas of process (see the appendices for a brief overview of Process Theory and the Process Algebra). The Buddhist theory of mind traces the path from sensory receptor through perception to interpretation and action. Its central concepts are the transient nature of psychological experiences, their conditionality, which in modern terms would be their contextual nature, and the emergent nature of the self (often referred to as "no mind"). These are remarkably modern ideas, compatible with Complex Systems Theory [2], Whitehead's Process Theory [3] and with modern neuroscience and psychology [4-8].

Various metaphors have been entertained over the centuries to aid in our understanding of the nervous system, usually grounded in the technology of the time: hydraulics (Hippocrates), electromechanics (Galvinism), mechanics (behaviour theory), computers (cognitive psychology), networks (functional connectivity). However, neuroscientists have discovered features of the dynamics of the nervous system which place it at odds with mechanistic conceptions and in keeping with process concepts. In particular there is [9-15]:

    1) the transient nature of neural activity,

2) the generative nature of neural activity,
3) the fungibility of the components which enter into any action,
4) the metastability of internal representations,
5) the contextuality of neural activity from perception to response,
6) the fundamental openness of the nervous system to the environment,
7) the importance of information (patterns, symbols, signs, semantics) rather than energy as a main determinant of activity
8) the absence of complete information and the importance of anticipation
9) the developmental origins of the components which form the nervous system,
10) the transient nature of neural connectivity,
11) the importance of embodiment,
12) the importance of extra-neural influences (somatic influences and gut microbiota),
13) the dependence upon multiple spatiotemporal scales,
14) the presence of a hierarchical organization with complex feedback relationships (bottom-up, top-down, and horizontal),
15) the fundamental stochastic nature of neural activity,
16) the causal relationship between hierarchical levels being that of emergence (vertical, horizontal, diagonal) rather than strict determinism or reductionism.

Taken together, these features demonstrate that the nervous system is not an "object" in the classical sense but rather forms a process a la Whitehead.

Rather than falling prey to "physics envy" and looking to mechanistic metaphors to understand the nervous system, it is necessary to turn to biological systems which could serve as metaphors, for it is only biological systems which possess the dynamical features noted above. Boldly, Rosen wrote [16] ". . . the basis on which theoretical physics has developed for the past three centuries is, in several crucial respects, too narrow and that, far from being universal, the conceptual foundation of what we presently call theoretical physics is still very special; indeed, far too much to accommodate organic phenomena (and much else besides)." (pg. 315)

Physicists have made effective use of two tools: formal mathematical modeling for representing, analyzing and predicting the behaviour of physical entities, and the reductionist paradigm, which assumes that the behaviour of composite or complicated entities can be understood as the sum of their parts. Mathematics, in its present form, is based upon notions of ideal, eternal, immutable, isolatable entities for which there is complete knowledge, at least in principle. Such notions are compatible with the physicist's conception of classical inanimate objects and the similarity is not accidental – mathematics and physics have grown together, mutually stimulating the growth of each other. The dynamical features of the nervous system have no counterpart in mathematics as it is currently formulated – there is a desperate need for collaborations between neuroscientists and mathematicians to develop mathematics which is appropriate for the description and analysis of process [13-15] (a brief overview of the Process Algebra, one such approach appears in the appendices). Moreover, reductionism as an explanatory paradigm fails – the presence of emergence, downward and horizontal causation, and contextuality all point to the necessity to approach the study of the nervous system simultaneously across multiple spatiotemporal scales.

A deep study of the dynamics of the nervous system requires the simultaneous observation and analysis of spatiotemporal scales ranging from neurotransmitters and receptors, cellular membranes, individual neural and glial cells, local cell assemblies, global cellular networks, somatic states, observable individual behaviour, individual subjective experience, dyadic relationships, familial relationships, social groups, communities, societies, cultures. The phenomena appearing at most of these scales are, for the most part, inaccessible to direct continuous observation, let alone simultaneous observation across scales. For the most part, researchers have tended to focus upon one spatiotemporal scale and attempted to explain its dynamics by reference only to that scale (everything else becomes initial or boundary condition, or noise). Such an approach will never tease out the relationships between the dynamics occurring at different scales. What is needed are simpler biological systems (proxies) which express dynamics across multiple spatiotemporal scales which is homologous to (not necessarily identical with) that exhibited by the nervous system. This would provide a biological model in which different hypotheses could be tested in a controlled manner, and links between dynamics at different spatiotemporal scales discerned, which could then be explored in the nervous system. Ideally such a biological system should be complex enough to exhibit homologies across a large range of spatiotemporal scales, yet small and simple enough to facilitate observation, data collection, experimental manipulation, and not run afoul of ethical considerations.

     Alternatives to the idea of the classical object have been proposed over the years. Anticipatory systems [17], Complex Adaptive Systems [2], Fractal Functionality and Functional Constructivism [14,15], and Naturally Occurring Computational Systems (NOCS) [12] are examples. The idea of a NOCS is to capture the formal dynamical features of behaving systems as they occur naturally in the real world. A NOCS is embedded within an active environment and its main goal is survival (classical objects do not need to survive as they are eternal). It is acted upon by and acts upon its environment. It possesses agency. It senses, interprets events, forms intentions, and acts. NOCS never possess complete information and seldom have the time needed to consider all possible options and select out only the best according to some arbitrary criterion (usually a fantasized utility function or fitness or truth valuation). NOCS utilize decision strategies that are *good enough*, that achieve some immediate ecological function or goal, in the moment, and then move on to the next task or challenge. Resilience, robustness and adaptability, are far more important than some ideal of optimality [12-15,18]. Collective intelligence systems (behaving systems composed of myriad agents interacting in the absence of a central authority) provide proto-typical examples of a NOCS. They possess a number of specific dynamical features [9-11]:
1) Self-Organization.
2) Stochastic determinism.
3) Interactive determinism.
4) Nonrepresentational contextual dependence.
5) Phase transitions, critical, and control parameters.
6) Broken ergodicity.
7) Broken symmetry.
8) Pattern isolation and reconfiguration.
9) Salience.
10) Irrelevance.
11) Compatibility and the mutual agreement principle [19].

which make them candidate proxies for nervous systems. The proto-typical collective intelligence system is the social insect colony.

The Social Insect Colony

A social insect colony, such as an ant colony, consists of a queen (primarily responsible for procreation) and a myriad of workers (specialized for various tasks such as brood care, midden work, patrolling, foraging, defense). All workers are female – males appear only when the colony needs to reproduce. Colonies range in size from a few hundred to as many as a million workers. Colonies are highly adapted to the environment within which they exist and the type of food that they rely upon. Colonies may reproduce once per season or continuously. They may remain localized in one area for decades or centuries, or move every few days. Nests may be large static structures, excavated underground or constructed from local materials above ground, or mobile, often formed of workers bodies. Colonies can distinguish between workers of their own colony and workers from different colonies, upon which they sometimes wage war. Some colonies make slaves of other species, or enter into symbiotic relationships, especially with certain plant species. Colonies exhibit diverse but ecologically salient responses to their environments. Individual workers possess brains formed of around 1 million cells. Individual workers are capable of making local decisions and interact with one another to effect decisions at the colony level. The activities of both individual workers and colonies appear to be stochastic, although structured, and there is no central authority overseeing or driving these activities [20-22]. Several years ago, it was suggested that the dynamical architecture of information processing by collective intelligence could serve as a model for the unconscious [23]. Here it is suggested that collective intelligence architectures might serve as proxy for studying the information dynamics of neural systems.

    A recent paper suggested the presence of deep dynamical homologies between collective intelligence systems and the neurobehavioural regulatory systems of the human brain, and that the study of collective intelligence systems, particularly social insect colonies, may shed light into neurodynamics [13]. This is not meant to imply that the dynamics of social insect colonies and nervous systems are identical, but rather that they utilize similar methods for representing, transferring and processing information, and that knowledge of information flow in social insect colonies might provide useful insights into how nervous systems process information. Social insect colonies are far more amenable to experimental observation and manipulation than is any human nervous system.

    One general reason to expect functional homologies between human nervous systems and social insect colonies is that both are examples of NOCS [12,18]. There are analogies between social insect colonies and organisms generally, expressed in the concept of the superorganism [20-22]. First suggested by Wheeler in 1911, the idea fell out of favor but has since seen a resurgence in recent years [22]. More particularly, the human nervous system and collective intelligence systems (and social insect colonies in particular) possess dynamics characterized by generativity, transience, emergence, contextuality, openness to the environment, stigmergy, creativity, and symmetry breaking [9-11], among many other properties. The nervous system has far more in common dynamically with collective intelligence systems then it does with hydraulic, mechanical or computer systems or networks.

    Functional homologies may be sought at one or more spatiotemporal levels. Here we consider homologies at three spatotemporal levels: cultural, psychological, and agent.

Cultural Level Homologies

At first glance it might seem odd to associate culture with social insects. Nevertheless, social insect colonies exhibit a vast diversity in terms of their adaption to their local environments [20-25]. Unlike humans, the origin of this diversity is likely genetic since there is no evidence for the passing of learned experience from one generation to another in social insect colonies. Still each species exhibits different social organizations, different responses to stressors and threats, all adapted to the local environment. Desert ants tend to live underground and forage the neighbouring region for food sources. These ants tend to be highly territorial and often there is war between adjacent colonies. Forest dwelling ants may form large mounds on the forest floor. They may form symbiotic relationships with nearby trees. Some ants engage in primitive forms of agriculture – cutting leaves and using them as a medium for growing fungi. Some species utilize slaves such as aphids. In response to the same alarm pheromone the workers of some species disperse while those of other species go deep within the nest, so that the response to the same sign acts as a species-specific signal eliciting distinct forms of response. This is similar to gestures which can have different meanings in different cultures.

**Psychological Level Homologies: Intransitivity in Decision Making**

As one example of the complexity of social insect colony decision making, consider the issue of intransitivity. The idea of an ideal mode of thought has its origins in philosophical and mathematical logic (where it serves to preserve the attribute of truth) and in economics (where rationality is a tool for enabling an agent to make choices leading to the achievement of an economic goal, such as maximizing a utility, maximizing a profit, or minimizing a cost or risk). One expression of rationality is in the formation of preference, which is important in an individual making one choice over another, and indifference, when a choice is made randomly with equal probability for the alternatives. An important feature of rationality in the determination of preferences is said to be the presence of transitivity [25]. This means that if an agent prefers choice A over choice B, and choice B over choice C, then they will prefer choice A over choice C, regardless of context, order of presentation or the presence of competing attributes. Preferences in which transitivity fails are said to be intransitive and characterized as being irrational or non-rational, and therefore to be avoided or dismissed. In both economic and psychological measurement theory, the existence of transitivity has the status of an axiom, since it is necessary in order that an ordinal scale exist for the observable being measured [25]. Since subjects are living agents and not machines, they are prone to errors and inconsistencies in their decision making. This leads to two forms of transitivity, a strong form (strong transitivity, ST) as depicted in the example above, and a stochastic form (weak stochastic transitivity, WST) where $x$ is preferred over $y$ if it is chosen more than 50% of the time [25].

There has been much debate as to whether the presence of intransitivity in human decision making represents some form of error which requires correction or represents a genuine form of decision making. Tversky [25] demonstrated the occurrence of consistent and predictable intransitivity in certain situations of decision making, for example choosing which of two alternatives to take a gamble on. Intransitivity has been observed in several studies [26,27]. Several authors have argued intransitivity need not be irrational [28] and that its appearance depends upon the context. Butler and Progrebna [29] point out that "Transitivity must hold either if a value attaches to each option without reference to other alternatives (choice-set independence), or if an equivalent value results after comparing and contrasting the attributes of the available choice options". Bar-Hillel and Margalit [30] presented three different contexts within which intransitivity might meaningfully occur 1) where intransitivity results from application of an ethical or moral choice rule; 2) where intransitivity results from application of an ethical or pragmatic choice rule; 3) where the choice is intrinsically comparative, depending upon multiple competing alternatives. In these contexts, intransitivity presents as a plausible consequence.

Some authors argue that transitivity is a universal phenomenon, and any deviation from transitivity is due to agent or experimental error [31]. There are criticisms of using pairwise comparisons to assess transitivity [32]. Errors leading to intransitivity include random preference, in which preferences are transient at each point in time but vary over time, and context-sensitive preference models, in which choice preferences are influenced by current and prior choice contexts [33].

Many authors have provided principled (most often mathematical) arguments for the existence of intransitivity. Fishburn [34] surveyed a number of models for intransitive preferences in diverse settings and suggested that transitivity is not essential to ensuring the existence of maximally preferred alternatives in a number of situations. People need not always engage in decision making that invokes transitivity, and reasonably so. Butler and Progrebna [29] conducted a set of lotteries and observed rates of transitivity and intransitivity, accounting for factors such as noisy variation. They found that "many typically transitive individuals are the same people who violate transitivity in the circumstances we identify. This suggests neither a transitive nor intransitive 'core' utility function can accurately describe preferences over all lotteries a person may encounter. …. in line with a constructed-preference paradigm". This experiment demonstrated that people use different strategies in different contexts; a strategy which appears rational in one context may not be in a different context. Far from being a liability, this makes human decision adaptable to different circumstances. There is no such thing as a one size fits all, or universal strategy. Klimenko [35] introduced a general measure of intransitivity, the evolutionary intransitivity parameter. He points out that "Human preferences that seem irrational from the perspective of the conventional utility theory, become perfectly logical in the intransitive and relativistic framework suggested here". He concludes that intransitivity should appear under any of the following conditions: relative comparison criteria, multiple incommensurable comparison criteria, multiple comparison criteria that are known approximately, comparisons of groups of comparable elements. An interesting formal analysis within the setting of game theory again shows that intransitivity need not be viewed as irrational, and indeed may sometimes be the preferred attribute of a decision strategy [36].

Decision making by individual workers and by colonies as a whole have been studied intensively. Workers of *Temnothorax albipennis* choose nest sites based upon several characteristics such as level of lighting (dark over bright) and entrance width (narrow over wide). Edward and Pratt [37] presented ants with two nest sites, A (dark with average entrance) and B (bright with narrow entrance). Workers choose either of these sites with equal likelihood. However, when suitable decoys are presented, the symmetry in preference is broken: with decoy A (dark with wide entrance) ants prefer A, while with decoy B (average light with narrow entrance) ants prefer B. This is a violation of rationality. This result was confirmed in a subsequent experiment [38].

Individual ants are also vulnerable to another class of contextual effects, namely contrast effects. Workers of *Lasius niger* appeared not to assess the value of food based on its ability to satiate them but rather compared to alternative food sources, even when the alternative food sources in question were not even present. Food received from other workers within the nest appeared to serve as a reference value [39]. Human judgments are frequently influenced by extraneous factors such as the presence of labels. Workers of *Lasius niger* show similar biasing by the presence of extraneous odors [40,41].

The most detailed study to date of collective decision making was conducted by Franks et al. [42]. They studied nest emigration by colonies of *Temnothorax albipennis*. They identified several attributes of potential nest sites which the ants appeared to utilize in making a decision including the brightness of the site, its height, and the width of the entrance and exposed the colony systematically to a range of paired alternatives. They showed that the colony appeared to express transitivity in its preference hierarchy and in addition, appeared to use a weighted additive strategy [42].

Edwards and Pratt [37] subjected the colony as a whole to the same set of alternatives and decoys that were presented to the individual workers. Unlike its workers, decisions at the colony level evaded falling prey to the decoy effect. However, the ability of the colony to utilize a rational strategy appears to depend upon contextual factors. For example, a study of foraging by *Myrmica rubra* [43] found that modifying the available choice set by increasing the number of nest entrances from one to two resulted in worse foraging outcomes.

In [44], researchers forced workers of *Temnothorax albipennis* to migrate from a high-quality nest and to choose between a mediocre and a poor nest site. The ants universally moved to the mediocre site. However, if ants were exposed to an alternative nest site prior to being forced to emigrate, and then later forced to choose between the familiar alternative and an unfamiliar alternative, they showed an aversion for the familiar alternative, even when that led to the choice of a poorer site. The intensity of this aversion was influenced not just by the quality of the alternative but also by the quality of the home nest. The authors used formal modeling to show that comparative strategies might manifest at the colony level, but individual workers could use absolute strategies combined with threshold-based decision rules, demonstrating how an experience-dependent, flexible strategy can emerge at the global level from a fixed-threshold strategy at the local level. In a subsequent study [45] workers were exposed to an alternative nest site of lower quality than their own and then forced to emigrate facing the familiar alternative and a novel alternative of similar quality. They avoided the familiar site and opted for the novel, breaking preference symmetry. However, if presented with familiar and novel high-quality sites, they maintained symmetry. The ability to assess and retain information about potential nest sites appeared due to the use of pheromones and landmark cues.

O'Shea-Wheller et al. [46] observed that individual workers appear to manifest a heterogeneous range of decision thresholds which manifest in the duration that they spend in a potential nest site and showed that the presence of heterogeneous thresholds allowed the colony to effect optimal, self-organized emigration decisions without the need for direct comparisons at the local level.

Doran et al. confirmed those results [47]. In their study of nest selection in ants they found the tendency of a colony to move was not based on the value of alternate sites in some abstract sense. Instead, colonies assessed nest-sites based on the potential fitness benefit of moving. Ultimately, because a nest site's value to an ant colony was context-dependent, two nests assessed under different conditions were not evaluated on the same transitive hierarchy. Furthermore, they were able to show that flexibility was not entirely relegated to the colony, but individual workers were also able to modify their response through changes in recruitment speed. Formal research has suggested that non-rational decision making may outperform rationality in certain contexts. Houston [48] used formal analysis of a mathematical model of foraging to show that the fitness value of any food item was contextual rather than absolute, dependent on its alternatives and its probability of being foraged. He argued that it was unlikely that natural selection could thus assign an absolute fitness value to each food option. Even if rationality was possible, it would perform sub optimally compared to context-dependent decision-making methods that violated a form of stochastic transitivity.

Sasaki et al [49] studied decision making by colonies of *Temnothorax rugatulus*. Through direct experimental observation and computational modeling, they studied the ability of colonies versus individuals to choose between nests having varying degrees of difference among them. Experimentally they showed that colonies outperform individuals when the degree of difference is small so that discrimination is difficult. When the degree of difference is large, and so discrimination is easy, individuals outperform colonies which are more prone to errors in such circumstances. They developed a computational model, which emphasizes the role of positive feedback (as does [50]). They showed that positive feedback enabled the colony to integrate information from individuals and enhance the discrimination between fine differences. However, when the differences are large, positive feedback can lock the colony into choices which ultimately turn out to be suboptimal.

**Agent Level Homologies**

In the nervous system the main agents undertaking information processing may be taken to be the neurons and glial cells. These cells are morphologically diverse and somewhat specialized to fulfill different roles and to secrete particular neurotransmitters. They are active cells – secreting chemical agents and in the case of neurons, generating electrical impulses, which enable them to act upon other cells. They are embedded in an active environment which contributes to both the maintenance and function of neurons. Cajal provided two postulates related to neuronal function: (1) the principle of dynamic polarization, which states that the electrical activity of the neuron flows from dendrites to the trigger zone of the axon; and (2) the principle of connectional specificity, which states that neurons do not connect purely at random but rather only to specific populations of neurons. Neurons thus have individual activity but also collective activity. There are many different theories concerning how the collective activity contributes to information processing: coherent oscillations [7,8], spike-train synchrony [7,8], TIGoRS [51,52], mass action [4] and chaotic dynamical phases [5] to name a few. The first two approaches are mechanistic, drawing upon analogies to signal processing. The latter three are complex systems approaches describing emergent effects. In all cases, neural dynamics exhibits a great deal of stochasticity, whether the release of neurotransmitters [53], tor the timing of spike trains [54].

Much is made of the organization of neurons into networks, but insufficient attention is paid to the fact that these networks, even if globally stable, are not locally stable. They are dynamic, especially at the dendritic and receptor levels. This enables otherwise spatially static neurons to explore a virtual space of neuronal dynamics (i.e. the space of maps describing the temporal dynamics of neurons). Indeed, every night, serotonin activity in the brain ceases, resulting in a contraction of dendritic spines which re-form again in differing geometries each morning [55].

The most described means of interaction between neurons is through the synapse. Typically the synapse is a specialized region linking an axon terminal to a dendritic spine. The axonal region (upstream neuron) is specialized to release vesicles containing neurotransmitters.

The dendritic region is also specialized, containing receptors which receive the neurotransmitter and, if the conditions are conducive, carry out some action.

Synapses come in several forms [7]:

1) Gap junction synapses, which are electrical and provide direct bi-directional interactions between neurons.

2) Synapses with ligand-gated receptors, which produce excitatory or inhibitory effects upon the local neuronal membrane when stimulated by an appropriate neurotransmitter.

3) Synapses with G-protein-coupled receptors, which serve a modulatory role, and which, upon receipt of a trans-threshold quantity of neurotransmitter, alter the response of the neuron to ligand-gated receptor activation by modulating various membrane parameters such as resistance, length and time constants, the duration of action potentials, and so on.

4) Synapses with tyrosine kinase receptors, which play a role in gene transcription and metabolic processes, and in turn play a role in disease induction as well as memory.

Two additional modes of interaction have been largely ignored. The first of these is ephaptic (or ephatic) transmission. This occurs when two axons are sufficiently close together so that electrical activity in one axon influences the electrical activity in the other. This is a form of neuronal transmission that is extremely difficult to study, but recent theoretical work suggests that it might play a greater role in the nervous system than previously recognized, at least in the dynamics of unmyelinated fibres [56, 57].

The second form is actually ubiquitous throughout the nervous system and likely plays a very important role in information processing and in modulating neuronal dynamics. This is volume transmission, originally associated with somatic influences on neural dynamics such as those arising from hormones, neuropeptides, cytokines, and interleukins, among many others. In recent years, however, it has been realized that neurons may also release large quantities of neurotransmitter directly into the extracellular space rather than delivering it to vesicles that participate in synaptic interactions. Volume transmission is thought to underlie phenomena such as mood, attention, arousal [58–60], and to regulate the dynamics of mesoscopic level cellular assemblies which operate over large spatio-temporal scales [60]. Somatic influences (including the effects of gut microbiota [61]) effected through the release of hormones, neuropeptides, and immune factors all modulate neurodynamics through volume transmission [7].

These different modes of interaction take place over a wide range of temporal scales: gap junction synapses are virtually instantaneous, ligand-gated synapses act over 0.3–5 milliseconds, G-protein-coupled synapses act over hundreds of milliseconds to minutes (as may volume transmission), whereas tyrosine kinase receptors have effects that can last up to weeks

The relationship between neuronal activity and behaviour is profoundly complex. It has long been known that behaviour is not reproduced as a fixed pattern elicited by some stimulus but, instead, is generated anew, "on the fly", each and every time it is elicited. Different neurons, different muscle fibres, different sequencing, and different timing all distinguish one instance of a specific behaviour from another, even if superficially these behaviours all appear to be the same or express the same functionality [9,13-15,62]. Moreover, so called internal "representations" of the external world such as internal maps and memories are metastable, meaning that different neurons are involved at different times, even though functionality at the macroscopic level appears to be preserved [63-65]. Neurons appear to be fungible.

Social insect colony workers are also morphologically diverse, forming distinct castes. In some species, morphology (caste) determines role [20,21]. In other species, workers may switch roles but there are rules governing which transitions are permitted, their duration, and whether they are reversible [24]. Workers interact using two modalities mostly: physical touch or chemical signals.

Physical interaction takes two general forms:

1) A worker may stroke or vibrate another worker, stimulating specialized sensory receptors

2) A worker may touch a worker using stereotyped movements, as in recruitment of tandem running, or sometimes simply pick up another worker entirely as in carrying.

Interactions through chemical mediators take three general forms [20-22]:

1) Workers may secrete, through specialized glands, chemical markers which remain on the surface of their bodies and which other workers access through touch, generally using their antennae. These chemicals may indicate affiliation, or specific needs in the case of pupae and larvae.

2) Workers may secrete chemicals into the larger environment, which then diffuse into large areas. These pheromones serve many purposes such as to mark trails, to signal danger and so on.

Workers may also interact with one another through physical changes in the environment. This is best observed in nest construction, in which the formation of certain patterned elements during the construction of the nest initiates a change in the rules by which workers act, causing them to create a new set of patterned elements, which may again induce further changes in rules. This phenomenon, the "incitement to work by the products of work" [20] is termed stigmergy.

**Functional Homologies between Neural and Collective Intelligence Systems**

In searching for functional homologies between neural and collective intelligence systems we are *not* seeking mechanical equivalence between these two systems. It is patently obvious that from a strictly mechanical point of view, the manner in which neural and collective intelligence systems facilitate interactions between their agents, neurons and workers, are physically, structurally, mechanically different, for the most part. We are interested in understanding how information is processed by both systems, and how information processing at the various spatiotemporal scales affects processing at the other scales. We are therefore interested primarily in the functional role that different modes of interaction play in each system, and to look for homologies between these functions, regardless of their mechanical implementation.

The following functional homologies seem plausible:

1) Agents can be separated into distinct categories by morphology/neurotransmitter in neural systems and by morphology/role in collective intelligence systems.

2) There is no central authority in either system. There is no plan, script, schema. Global decisions appear to be collective, though individual agents make their own decisions. The global decision arises through a collective mechanism – for neurons, mass action [4], for workers, quorum thresholds [42, 46, 66,67].

3) In both systems, stochasticity plays a large role.

4) In both systems, global behaviours are generated "on the fly"; participating agents are recruited in the moment and are, to varying degrees, fungible.

5) In both systems, global behaviours are emergent and dependent upon interactions among the agents.

6) In both systems, interaction patterns conform to a network, but, to varying degrees these networks are dynamic, changing over different time scales.

7) Both systems are open systems – they require ongoing interaction with an environment, not merely for survival, but for the contexts and conditions which stimulate and shape global behaviour.

8) Both systems are active – they act upon their environments, modifying them in ways which create new contexts and conditions, thus altering global behaviour.

9) In both systems there are two principal environments – there is a local environment, body for neural systems and nest for collective intelligence systems, which constitute emergent organisms in their own right, and there is a global environment, which for both consists of the external world. Both systems maintain the functional integrity of the local environment while interacting with the global environment for resources and defense.

10) Both systems utilize physical local means to mediate interactions among agents – neurons use gap junctions and ephaptic transmission, workers use vibration and touch.

11) Both systems utilize local chemical means to mediate interactions among agents – neurons use ligand-gated, G-protein coupled and kinase receptors, workers use surface secretions.

12) Both systems utilize global chemical means to modulate interactions among agents – neurons use volume transmission, workers use diffusing pheromones

13) In both systems, feedback from both local and global environments involves both local and global aspects. For neurons, there is direct local feedback from somatic receptors and global feedback via circulating hormones, neurohumoral and neuroimmunohumoral factors. For workers, there is local feedback from stigmergic properties of the physical nest, and global feedback from broos pheromones. In the global environment there are also local and global aspects depending upon whether the interaction requires direct contact (local) or occurs through some diffusive, dispersive or radiating intermediary (global).

One major difference between neurobehavioural regulatory systems and collective intelligence systems is that the agents of a collective intelligence are generally free to physically move through their environment, whereas the neurons in a neurobehavioural regulatory system are fixed in space. However, as noted above, the connectivity among neurons is functional more than it is structural; it is dynamic and allows neurons to explore a virtual space of connectivity, and thus of dynamics, which may as functionally important as a worker's ability to explore a physical space.

As can be seen, there are a very large number of functional homologies between neural and collective intelligence systems. Although mechanisms may differ, these similarities in function suggest that the induction of systematic changes in function in a collective intelligence system could provide insights into how a similar alteration of function might affect information processing within neural systems. In this way it might be possible to study systematic, generic features of information flow and its associated functionalities. In situations in which a particular functionality depends upon the action of a specific mechanism, and is uniquely linked to that mechanism, then targeted studies directed at the specific mechanism in that specific system could be carried out. The ability to determine when such specific studies need to be carried out rather than relying on generic results might still result in a significant saving in resources, and ultimately in the lives of organisms. Moreover, as noted above, it is potentially much easier to simultaneously study multiple spatiotemporal scales in collective intelligence systems, and to carry out modifications of such systems to examine the roles that different functionalities play in information flow, than it is with neural systems. The discovery of generic results concerning the dynamical linkages between different spatiotemporal scales would be a vast improvement over our current situation.

The study of collective intelligence systems is not meant to supplant investigations into neural systems but rather to provide insights which might help guide future research. The cultivation of collective intelligence systems, particularly social insect colonies such as ant colonies, is relatively easy and inexpensive and their use as experimental subjects, especially for examining interactions across multiple spatiotemporal scales, is more practical and less prone to provoke ethical dilemmas, than is the case for neural systems. less likely to cause ethical issues, and due to their smaller size, are more amenable to detailed mathematical modeling, some of which has already been carried out.

**Grand Unified Theories: A Comment**

Physicists have long sought the grand theory of everything, the theory which would describe all of the forces and fundamental entities of nature and, at least according to the reductionist paradigm, from which everything else could be derived and its behaviour predicted. So far, they have not succeeded. The arrival of complexity science should have dealt a death bow to ideas of reductionism, and this is mostly true, but the idea still shows up in some disciplines. Similar dreams have appeared from time to time in psychology and neural science although they seem out of favour at the present time. This rise of network and information science has also come with expressions of general principles such as the Free Energy Principle, or the ubiquity of small world networks or fractals as having some sort of causal significance.

The approach advocated here is somewhat more modest. The idea of the Process Algebra comes from the recognition that the types of dynamics observed at the various spatiotemporal scales in neural systems and collective intelligence systems all share certain common features. These features can be nicely summarized within the concept of process. The Process Algebra is meant to provide a mathematical language for describing the actions and interactions of processes based on Whitehead's Process Theory, which in turn was meant to capture fundamental properties of organisms. The Process Algebra itself is not a theory. It is a high-level mathematical language for describing processes and within which theories of processes can be situated. Its advantage is that it provides a single, unifying mathematical language for describing processes across the various spatiotemporal scales. This makes it easier to compare functionalities at these different scales and to describe their interactions.

The dynamics at a specific spatiotemporal scale may be thought of as unfolding according to its own set of rules or principles or laws, dispositions or imperatives. Functionalities appearing at larger and smaller spatiotemporal scales can then be viewed as contexts for the dynamics at the level of interest. Lower-level functionalities may also be viewed as setting dispositions, establishing initial conditions, and as providing sources of intrinsic noise. Higher level functionalities may be though of as providing stimuli, boundary conditions or constraints, and as sources of extrinsic noise. Such comparisons are not exclusive, and a great deal of overlap may occur. Generally, dynamics at one level is thought of as emergent upon the dynamics taking place at lower levels subject to constraints imposed by dynamics at higher levels.

If a conjecture concerning general principles is permitted, one suggestion is that of relative dynamical autonomy. In a strongly emergent system such as a neural system or collective intelligence system, the dynamics at each levels acts, at least over relatively short time scales, to preserve its coherence – its rules, principles, and laws. Influences from lower and higher levels may be treated as noise. The dynamics at each level, therefore, attempts to minimize the impact of these perturbations on the subsequent behaviour of the system. However, over extended periods of time, the presence of repeated or persistent fluctuations, perturbations or stimuli, or changes in contexts, may induce a gradual drift in the rules, principles or laws present at the level of interest, subject to the constraints imposed by the contexts. Adaptation may be thought of as minimizing the effect of long-term perturbations by altering the rules. That is, in the short term, the dynamics at each level may be viewed as stable, and error correcting (or homeostatic), while over the long term the dynamics at each level may be viewed as adaptive, subject to constraints imposed by the contexts.

Conclusion

Based upon this brief review of functional homologies between neural and collective intelligence systems, it is argued that collective intelligence systems, particularly social insect colonies, provide a proxy for studying generic features of information flow and processing, for studying the roles played by different functionalities, and for examining how information flow is affected by interactions among the various spatiotemporal scales among which these functional systems are distributed. Social insect colonies, in particular, are easy to cultivate and maintain. Their relatively small size and accessibility across multiple spatiotemporal scales makes them ideally suited for experimental manipulation and formal theoretical and mathematical modeling. One language for theoretical description and analysis is the Process Algebra, some details of which are briefly presented.

Appendix A
Appendix A.1. Process Theory
The core ideas of Whitehead's Process Theory [3] have been presented elsewhere [68] but are summarized here concisely for reference. Described as a philosophy of organism, Whitehead proposed a metaphysics in which reality consisted of an ever-changing flux of phenomena organized into coherence by some form of underlying subjectivity. The subjectivity in Whitehead's theory is termed "prehension", which loosely refers to the incorporation of prior information into the newly emerging elements of reality.

Whitehead called these most primitive elements of reality "actual occasions". According to Whitehead, a process consists of a sequence of events having a coherent temporal structure in which relationships between events are considered more fundamental than the events themselves. Becoming is a fundamental aspect of process, whereas being and substance arise from the actions of process. In process theory, entities are considered to be generated as opposed to simply existing. Whitehead's actual occasions are transient entities: they come into being, exist long enough to pass on whatever information they represent, and then fade away. There is a subjective, meaning-laden thread linking these events. The entities that make up our observable reality are emergent from these actual occasions.

Processes generate the actual occasions that constitute space–time, and as such may be considered to exist outside of space and time. The idea that (quantum) phenomena might possess features which exist outside of space–time has been suggested by Bancal and Gisin and colleagues [69], and by Aerts and Sozzo and their colleagues [70]. They can exist either in a state of activity, in which they generate actual occasions, or they can exist in a state of inactivity, in which they are merely potentialities for a future state of activity. Transitions between these states depend upon the flux of actual occasions in the moment and the interactions among the currently active processes.

Alterations in the characteristics of processes occur through interactions among processes, dependent upon their compatibility [19] and triggered by the appearance of specific actual occasions.

The formal representation of the process in process algebra has been described in detail elsewhere [71-75], and so will merely be summarized here. Process algebra rests upon the realization that mathematical structures can be *generated* with the use of combinatorial games (in particular, Ehrenfeucht–Fraïssé games) [76-78], together with the fact that many basis sets in functional analysis, such as the Hilbert space of NRQM, are reproducing kernel Hilbert spaces [79]. Given a reproducing kernel Hilbert space $H(X)$ with base space $X$, one can find a discrete subspace $Y$ of $X$ (sampling subspace), and a Hilbert space $H(Y)$ on $Y$, such that each function in $H(Y)$ can be lifted to a function in $H(X)$ via interpolation. Interpolation means that if $\Psi(z)$ is a function in $H(X)$, then for each $y \in Y$ there exists an interpolation function $\Psi_y(z)$ on $H(X)$, such that $\Psi(z) = \sum_{y \in Y} \Psi(y) \Psi_y(z)$.

One can use either Whittaker–Shannon–Kotel'nikov sinc interpolation theory (for $Y$ being a discrete lattice), or Fechtinger–Gröchenik interpolation theory may be used instead [79] (for non-uniform spaces satisfying the Beurling density [80]). Processes can be modeled heuristically as (epistemologically equivalent [71-75]) combinatorial games, which generates a discrete space of primitive events from which the larger events emerge via interpolation. The discrete subsets $Y$ are considered to be fundamental, $X$ is an *interpretation* selected by an observer, the elements of $H(Y)$ are the ontological state (wave) functions, and the elements of $H(X)$ are derived (emergent) through an (arbitrary) interpolation procedure. The important point is that the elements of the space $Y$ are created in distinct generations, and the value of the function at each point is determined by propagating information from prior elements by means of a causal propagator, $K$. Thus, $\Psi(y) = \Sigma_i K(y,i) \Psi(i)$, where the sum is over immediately prior elements i. Probabilities are emergent, arising from interactions between processes.

The discrete subsets are called *causal tapestries* and their individual points are called *informons*. A detailed description of informons is given in [71-75], and the interested reader is referred there because the details are not essential to the discussion in this paper.

A fundamental tenet of this model is that a process does not change state unless in interaction with other compatible processes. The concept of compatibility between interacting complex systems was first proposed by Trofimova [19]. Compatibility $\Xi(\mathbf{P},\mathbf{M})$ is conjectured to be a function of fixed factors (e.g., mass, charge, coupling constants) and of the local compatibilities. The probability of an interaction taking place $\Pi(\mathbf{P},\mathbf{M})$ is conjectured to be a function of the compatibility, $\Pi(\mathbf{P},\mathbf{M}) = \chi(\Xi(\mathbf{P},\mathbf{M}))$. The precise form of these functions depends upon the particular case, but it can be expected to depend, in part, on the local process strength $\Psi^*(y)\Psi(y)$.

Appendix A.2. Process Algebra

Process algebra is the formal language for describing interactions between processes.

There are additional technical aspects such as the process covering map and the configuration space covering map, whose details can be found elsewhere [71-75]

An important concept is that of *epistemological equivalence*. Epistemological equivalence of two processes, **P** and **Q**, means that their global Hilbert space interpretations, $\Psi^P(z)$, $\Psi^Q(z)$, respectively, are equal, i.e., $\Psi^P(z) = \Psi^Q(z)$.

If two processes are epistemologically equivalent, then the specifics of generation do not matter. They generate the same emergent state functions and therefore will yield the same predictions. This is useful because processes can be modeled heuristically based upon mathematical convenience, just so long as they are epistemologically equivalent to any real processes. In particular, one can use processes based upon combinatorial games which have particularly valuable characteristics [76-78].

Processes may influence one another in two different ways. The first (*coupling*) involves the generation of individual informons, their relative timing, as well as the sources of information which enter their generation. Coupling results in epistemologically equivalent processes; thus, properties are unaltered. The second (*interaction*) involves the activation or inactivation of individual processes and the creation of new processes. Epistemological equivalence is broken, and properties are altered.

Two processes, $P_1$, $P_2$, may be independent, meaning that neither constrains the actions of the other in any way. This relationship is denoted simply by the comma ",". Compound processes (R > 1) can be formed from primitive processes (R = 1) by various coupling operations. A coupling affects timing and information flow. Two processes may generate informons concurrently (*products*) during each round, or sequentially (*sums*), with only one process generating informons during a given round. Information from either or both processes may enter into the generation of a given informon (free), or information incorporated into an informon by a process may only come from informons previously generated by that process (*exclusive*). This leads to four possible operators:

1 Free sequential (free sum): $P_1 \hat{\oplus} P_2$;

2. Exclusive sequential (exclusive sum): $P_1 \oplus P_2$;

3. Free concurrent (free product): $P_1 \hat{\otimes} P_2$;

4. Exclusive concurrent (exclusive product): $P_1 \otimes P_2$.

The operation of concatenation is used to denote processes that act in successive generation cycles. Thus, $P_1 \cdot P_2$ (or simply $P_1 P_2$) indicates that $P_1$ acts during the first generation cycle, whereas $P_2$ acts during the second generation cycle.

Interactions break epistemological equivalence, and can do so in myriad ways. Interactions between processes may activate an inactive process or inactivate an active process. In addition, an interaction among processes $P_1$, $P_2$, ..., $P_n$ may generate a new process, **P**, which can be described in functional form as $F(P_1, P_2, \ldots, P_n) = P$. If $\Theta(P_1, P_2, \ldots, P_n)$ describes a coupling among $P_1, P_2, \ldots, P_n$, then the functional relationship may be described using the operation of concatenation, as $\Theta(P_1, P_2, \ldots, P_n)\,P$.

There are potentially so many different types of interactions; therefore, a set of generic operators comparable to those above are used to indicate the presence of an interaction with the specifics to be spelled out if known. Thus, there are:

1. Free sequential (free interactive sum);
2. Exclusive sequential (exclusive interactive sum);
3. Free concurrent (free interactive product);
4. Exclusive concurrent (exclusive product).

Independence, sums, and products are commutative, associative, and distributive operations. Concatenation is non-commutative and non-associative in general. The zero process, **O**, is the process that does nothing.

The basic rules for applying these operations in combining processes are the following:

(1) The free sum is only used for single systems and combining states which possess identical property sets (pure states);

(2) The exclusive sum is used for single systems and combining states which possess distinct property sets (mixed states);

(3) The free product is used for multiple systems which possess distinct characters (scalar, spinorial, vectorial, and so on) (for example, coupling a boson and a fermion);

(4) The exclusive product is used for multiple systems which possess the same character (for example, coupling two bosons or two fermions).